\documentclass[preprint]{aastex}

\shortauthors{Manset, Bastien \& Bertout}
\shorttitle{Orbit-induced polarimetric observations of AK~Sco}

\begin{document}

\title{Polarimetric variations of binary stars. VI. Orbit-induced
variations in the pre-main-sequence binary AK Sco\altaffilmark{1}}
\author{N. Manset}
\affil{Canada-France-Hawaii Telescope Corporation,
65-1238 Mamalahoa Hwy, Kamuela, HI 96743, USA}
\email{manset@cfht.hawaii.edu}
\author{P. Bastien}
\affil{D\'epartement de Physique, Universit\'e de Montr\'eal,
C.P. 6128, Succursale Centre-Ville, Montr\'eal, QC, H3C 3J7, Canada, and
Observatoire du Mont M\'egantic}
\email{bastien@astro.umontreal.ca}
\author{C. Bertout}
\affil{Institut d'Astrophysique de Paris, 98 bis bd Arago,
75014 Paris, France}
\email{claude.bertout@obspm.fr}

\altaffiltext{1}{Based on observations made with ESO Telescopes at the
La Silla Observatory.}

\begin{abstract}
We present simultaneous $UBV$ polarimetric and photometric observations
of the pre-main-sequence binary AK~Sco, obtained over 12 nights,
slightly less than the orbital period of 13.6 days. The polarization is
a sum of interstellar and intrinsic polarization, with a significant
intrinsic polarization of 1\% at 5250\AA, indicating the presence of
circumstellar matter distributed in an asymmetric geometry. The
polarization and its position angle are clearly variable on time scales
of hours and nights, in all 3 wavelengths, with a behavior related to
the orbital motion. The variations have the highest amplitudes seen so
far for pre-main-sequence binaries ($\approx$1\%, $\approx30$\arcdeg)
and are sinusoidal with periods similar to the orbital period and half
of it. The polarization variations are generally correlated with the
photometric ones: when the star gets fainter, it also gets redder and
its polarization increases. The color-magnitude diagram $B-V$, $V$
exhibits a ratio of total to selective absorption $R=4.3$ higher than in
normal interstellar clouds ($R=3.1$).  The interpretation of the
simultaneous photometric and polarimetric observations is that a cloud
of circumstellar matter passes in front of the star, decreasing the
amount of direct, unpolarized light, and hence increasing the
contribution of scattered (blue) light. We show that the large amplitude
of the polarization variations can not be reproduced with a single
scattering model and axially symmetric circumbinary or circumstellar
disks.
\end{abstract}


\section{Introduction}
AK~Sco (HD~152404 = IRAS~16514-3648 = HBC~271 = HIP~82747) is a
double-lined spectroscopic binary with a period of 13.6093~d, an
eccentricity of 0.469 (Mathieu 1994), and a projected separation of
0.143 AU (Jensen et al. 1996a). At a distance of $145^{+39}_{-25}$~pc
(Bertout, Robichon, \& Arenou 1999), it might be an outlying member of
the Upper Scorpio subgroup of the Sco-Cen association (Andersen et
al. 1989).

AK~Sco is estimated to be $6\times10^6$ yr old (Andersen et al. 1989)
and has strong Li absorption lines (Herbig \& Rao 1972). However, its
classification is unclear. The NIR excess and large irregular light
variations are more typical of a classical T~Tauri star (CTTS), but the
weak emission lines make it a weak-line T~Tauri star (WTTS) (Andersen et
al. 1989). The spectral type is F5 V (Herbig \& Rao 1972, Hamann \&
Persson 1992) which puts it in a group intermediate between T~Tauris
stars and F-type stars according to Th\'e, de Winter, \& P\'erez (1994).

AK~Sco has been known for decades to be highly variable at optical and
IR wavelengths. Periods of roughly constant optical brightness at an
average level are interrupted at irregular intervals by rapid variations
(Bibo \& Th\'e 1991), with no correlation (at least permanent) with
orbital period (see Jensen \& Mathieu 1997). The large photometric
variations are attributed to variable obscuration by circumstellar (CS)
dust condensations (Andersen et al. 1989, Hutchinson et al. 1994).

The submm continuum emission must arise outside the binary orbit, in a
circumbinary (CB) disk (Jensen \& Mathieu 1997) with a predicted
dynamically-cleared gap between 0.032 and 0.48~AU. This gap can
reproduce the observations if it is filled with optically thin material
to reproduce a strong 10 \micron\ silicate feature. However, since there
is no evidence for a NIR deficit of emission, this gap is not absolutely
necessary (Jensen \& Mathieu 1997).

Numerical simulations of the evolution of circumbinary disks by
G\"unther \& Kley (2002) show that in the dynamically-cleared gap in the
CB disk, spiral arms bring CS matter from the CB disk to the CS disks
around each star, in agreement with Jensen \& Mathieu's (1997)
interpretation of the SED with a gap filled with optically thin
material. Accretion is predicted to be synchronized with apastron, but
AK~Sco does not exhibit clear periodic excesses of luminosity at those
phases.

A model matching spectroscopic and photometric observations includes two
identical stars ($M = 1.5M_{\odot}$, $R = 2.2R_{\odot}$, $L =
0.9L_{\odot}$) with an orbital inclination of $\sim$63\arcdeg, embedded
in a dense cloud with fine structure near the stars and grains larger
than those of standard interstellar clouds. No uncertainty is given for
the orbital inclination, but eclipses, which have not been observed so
far, would be expected if $i \ga 70$\arcdeg. The cloud has a dimension
similar to the size of the orbit and was determined to be the cause of
the irregular variations. A cooler dust component also exists at a
larger distance at $\approx 10$~AU (Andersen et al. 1989).

AK~Sco's SED was also fit by Gregorio-Hetem \& Hetem (2002) to give
stellar radii of 1.69$R_{\odot}$, a disk radius of 10~AU and an
envelope radius of 1600~AU. The optical depth was $\tau=0.38$ and the
inclination 61\arcdeg. The authors also classify AK~Sco as a young main
sequence star.

Alencar et al. (2003) have also used photometric and spectroscopic
observations to find the physical parameters for the two near-identical
stars: $M = 1.35\pm0.07M_{\odot}$, $R = 1.59 \pm 0.35 R_{\odot}$, with
an orbital inclination 65\arcdeg $< i < $ 70\arcdeg. The dust
obscuration was also computed and reveals the existence of substructure
at a scale of a stellar diameter. One side of the orbit would also be
more obscured than the other.

Polarimetry of binaries, spectroscopic or visual, can be very useful to
learn about the geometry of their systems (see the recent review by
Manset 2004, in preparation).  In a recent series of papers (Manset \&
Bastien 2000, 2001; hereafter Papers I and II), we have studied
numerically the polarization variations produced by electrons and dusty
envelopes surrounding binary stars, and we have also developed tools to
compare observations of pre-main sequence (PMS) binaries with analytical
and numerical models and applied them to about two dozens PMS binaries
(Manset \& Bastien 2001b, 2002, 2003; hereafter Papers III, IV, and V).

In this paper, we present simultaneous polarimetric and
photometric observations of AK Sco (Sec. 2), use the
tools developed earlier to analyze these data (Sec. 3),
and compare with the photometric data (Sec. 4).
Finally, we compare our results with those of other
PMS binaries (Sec. 5) and with numerical simulations
(Sec. 6).

\section{Observations}
The photometric and polarimetric observations were carried out between
1982 February 12$-$24 with the European Southern Observatory (ESO)
photometer and polarimeter mounted respectively on the 50-cm and 1.0-m
telescopes at La Silla, Chile. Standard $UBV$ bandpasses were used for
the photometric observations. The polarimetric bandpasses approximate
the usual $UBV$ bands, and are defined by the following central
wavelengths and bandwidths: 3550 (700), 4300 (1000), and 5250 (500)
\AA. For the polarimetry, a 14\arcsec\ aperture hole was used. Data were
reduced using the ESO data reduction programs.

Primary photometric standard stars and two comparison stars within
1\arcdeg\ from AK~Sco, HR~8331 (SAO~208130, K0) and HR~8351 (SAO~208088,
G5), were observed on every night during the observing period, except on
1982 February 14-15 (UT). Table~\ref{Tab-compstars} gives the average
values, standard deviations, and the number of observations for the two
comparison stars.  These standard deviations are somewhat larger than
expected for a good photometric site such as La Silla, because the stars
(comparison stars and AK~Sco) have purposely been observed starting at
large air masses in order to increase the time coverage as much as
possible. This has no significant effect on the results because the
observed variations are much larger than the errors. The standard
deviations are larger in the $V$ band (and also B and U, although not
given) than in the colors because the atmospheric fluctuations are
larger in the total extinction (e.g., $k_{V}$ for $V$) than they are in
the colors (a differential effect, e.g., $k_{B-V}$ for
$B-V$). Nevertheless, the two comparison stars clearly show a stable
non-variable behavior, which demonstrates the photometric variability of
AK~Sco. The photometric data are listed in Table~\ref{Tab-phot}.

Standard polarized and unpolarized stars have been observed throughout
the observing run. The instrumental polarization was found to be $<$
0.05\% and was therefore neglected. The polarization data are given in
Tables~\ref{Tab-data3550} to \ref{Tab-data5250}.

Large amplitude variations in both photometric and polarimetric data are
readily noticed. These amplitudes are typically 0.5~mag in $V$, 0.12~mag
in $B-V$, and 0.20~mag in $U-B$, and 0.9\%, 47\arcdeg\ at 3550\AA, 1.0\%,
30\arcdeg\ at 4300\AA, and 0.8\%, 22\arcdeg\ at 5250\AA. The most
striking feature is that usually when the star is bright, the
polarization is small and when the polarization is large, the star is
fainter and also redder.

In the next section, we first estimate the contribution of interstellar
polarization to the observed polarization, then discuss the variability
and periodicity of the polarization.

\section{Polarimetry}

\subsection{Estimate of the interstellar and intrinsic polarization}
The Heiles (2000) polarization catalog, which contains polarimetric
observations for more than 9000 stars, was used to estimate the
interstellar (IS) polarization in the vicinity of AK~Sco by selecting 23
stars located within 80~pc and 15\arcdeg\ of it. The map of IS
polarization is presented in Figure~\ref{Fig-PolIS_AKSco}. The
IS distance-weighted polarization calculated by averaging the
polarization of the 23 stars is 0.70\% $\pm$ 0.16\% in the $V$ band. The
IS grains alignment is relatively good because the non-weighted average
position angle is $19\pm9$\arcdeg, whereas the distance-weighted value
is $165\pm5$\arcdeg. See Paper~V for more details about our averaging
procedure. 

AK~Sco's average polarization ($N$=27) is 0.70\% at 132\arcdeg\ at
3550\AA, 0.84\% at 132\arcdeg\ at 4300\AA, and 1.0\% at 133\arcdeg\ at
5250\AA\  (see Tables~\ref{Tab-data3550} to \ref{Tab-data5250} for the
detailed observations). These levels of polarization are comparable to
the IS polarization and indicate the presence of IS polarization in
AK~Sco's measurements. However, the weighted and non-weighted IS
position angles are different than AK~Sco's position angle values, and
indicate the presence of intrinsic polarization. We therefore conclude
that AK~Sco's observed polarization is a sum of IS and intrinsic
polarization. Bastien (1985) had also found that AK~Sco's polarization
is in part intrinsic and in part interstellar.

Using our estimate of the IS polarization, 0.70\% at 165\arcdeg\ in the
$V$ band, and Serkowski's law for IS polarization (Serkowski, Mathewson,
\& Ford 1975), we have found the IS polarization value for the 3
observed wavelengths. We have assumed that $\lambda_{\rm
max}=5500\;$\AA\ for Serkowski's law (a typical value) and that the
position angle of the IS polarization is constant as a function of
wavelength. After subtracting this estimated IS polarization from the
observed one, we get the estimated intrinsic polarization for AK~Sco:
0.70\% at 108\arcdeg\ at 3550\AA, 0.83\% at 109\arcdeg\ at 4300\AA, and
0.94\% at 112\arcdeg\ at 5250\AA. AK~Sco's intrinsic polarization is
high. Data taken for other PMS binaries (but at a redder wavelength,
7660\AA, see Papers~III $-$ V) indicate that AK~Sco has one of the
highest intrinsic polarization.

The significant intrinsic polarization indicates that the circumstellar
or circumbinary material around this binary has an asymmetric
configuration, for example, a disk rather than a spherical shell.

Two other facts point to an intrinsic polarization component in the
observed polarization for AK~Sco. First, AK~Sco shows clear periodic
variations (see below) that cannot be of IS origin. Second, the position
angle of the polarization varies as a function of wavelength (see
below). Observations of polarized standard stars (whose polarization is
of IS origin) by Schmidt, Elston, \& Lupie (1992) show that the maximum
observed rotation of the position angle between 4350 and 8580\AA\ is
$\approx 0.25 \arcdeg$, much less than the $\sim 10 \arcdeg$ rotation
seen for AK~Sco.

As discussed by Dolan \& Tapia (Dolan \& Tapia 1986 and references
cited), a wavelength-dependent position angle could in theory be
attributed to multiple IS clouds containing different grain sizes
magnetically aligned in different directions, or to an IS cloud where
there is a continuous rotation of the grains' orientation, but, once
again, the variability argument points to the presence of intrinsic
polarization.

\subsection{Variability analysis and comparison with previous data}
Already from the data presented in the tables and figures, it is obvious
that AK~Sco's polarization is variable, as had been found by Serkowski
(1969), Bastien (1985), and Drissen et al. (1989). Nonetheless, we
applied various tests to check the polarimetric variability or stability
of AK~Sco: minimum and maximum values, variance test, $Z$ test,
and finally, a $\chi^2$ test. It should be noted that these tests
usually assume that the parent distribution of the quantity measured
(here, $P$, $\theta$, or the Stokes parameters $Q$ and $U$) is
Normal. Since $P$ and $\theta$ are not distributed Normally (Serkowski
1958), these tests should in general be applied only to the Stokes
parameters $Q$ and $U$.

We have calculated the variance of the sample $\sigma_{\rm sample}$ and
compared it to the standard deviation of the mean $\sigma_{\rm
mean}$. For a set of observations of a non-variable quantity, the sample
variance will be small and similar to the standard deviation of the
mean. But if there is variability, the ``width'' of the observations, or
variance, will be greater than the standard deviation of the mean.

For the $Z$ test, if the data are ``well behaved'', or ``consistent'',
$Z\approx1$ within its standard error. If $Z$ differs significantly from
1, then there may be variability. Lastly, $\chi^2$ values are
calculated for $Q$ and $U$ separately, using 1$\sigma_i$ and
$1.5\sigma_i$. The probability to obtain a given value of $\chi^2$ in a
Gaussian distribution is found for each of the four $\chi^2$ values. The
star is variable if at least 2 of the four $\chi^2$ values are over
0.95; the star is suspected to be variable if one out of four $\chi^2$
values is over 0.95.

Details and formulas for these tests were presented in a previous paper
(Paper~IV). Results of the tests are presented in Table~\ref{Tab-Var}
and clearly indicate that AK~Sco's polarization was variable for all 3
filters, over a 12-day period in February 1982. Variations are observed
from night to night and also within the same night, and this variability
is more pronounced at 4300\AA.

AK~Sco presents one atypical observation, taken on February 13 in the
$U$ filter, where the position angle measured departs markedly from the
other values (see Table~\ref{Tab-data3550} and
Figure~\ref{Fig-3550all}). Atypical observations such as this one were
also found in many of the PMS binaries studied previously (Papers~III,
IV, and V) and are not related to the orbital motion but probably to
eruptive-like events. Since the data for AK~Sco was obtained with an
instrument different than the one used for Papers~III$-$V, this confirms
that it is not an instrumental effect. 

Since the position angle varies, it implies that if the polarization is
due to a cloud in orbit around the stars, the line of sight is not in
the orbital plane, in agreement with the orbital inclination found by
others.

Serkowski (1969) observed AK~Sco with $U$, $B$, and $V$ filters and a
25\arcsec\ aperture hole over a 454 day period. Although the
uncertainties are rather high (mean error 0.11\% in the $V$ and $B$
filters, higher in the $U$ filter), the data indicate variability on
time scales of days in all three filters. The polarization changes from
0.3 to 0.8\%, and the position angle, from 60\arcdeg\ to
145\arcdeg. Interestingly, the polarization variations sometimes go in
opposite directions in the $V$ and $B$ filters; when the polarization in
the $V$ filter increases, the polarization in the $B$ filter
decreases. The variations in position angle usually, but not always,
increase or decrease at the same time in both filters. There are no
long-term trends (the mean polarization does not change much from month
to month or over a few months). The mean position angle is the same in
the $B$ and $V$ filters.

Bastien (1985) has observed AK~Sco in various filters. As with the
Serkowski (1969) data, these observations show variability on time
scales of a day, with variations in all filters, and up to 1\% and up to
45\arcdeg. The amplitude of the polarimetric variations is greater in
the $U$ band than in the visual bands. On one occasion, the polarization
decreased in the $U$ and $V$ filters while increasing in the $B$
filter. The average polarization in all wavelengths was significantly
higher during Bastien's observations (in 1981) that during Serkowski's
(from 1967 to 1969).

Drissen et al. (1989) have observed AK~Sco at 7500\AA\  in 1986. The data
indicate once again variability on a nightly basis. The average
polarization seems to be at the same level as that of 1981. 

Hutchinson et al. (1994) observed AK~Sco in $B$ and $I_c$ in 1986. The
level of polarization and the position angle are similar to the 1980's
data. They noted that the flux in the infrared $N$ band rises as the
visual flux declines, which means the $V$ band variations seen during
their run are unlikely to be caused by the binarity. They also concluded
that the occultation model is consistent with the data in some cases but
does not always work with the simultaneous observations in the optical
and IR.

All these data indicate that polarimetric variability is mostly present
on time scales of hours and days, and decades (when comparing
Serkowski's data to all the other sets of data). Variations are observed
at all wavelengths. The variations for different filters usually go in
the same direction; 2 different sets of data have shown variations in
opposite direction for the $V$ and $B$ filters. The amplitude of the
variations is higher in the blue than in the yellow. No variation that
could mimic the behavior of the polarization during an eclipse was
observed.

\subsection{Periodic polarimetric variations}
Figures~\ref{Fig-3550all} to \ref{Fig-5250all} present the polarimetric
observations $P$ and $\theta$, and the Stokes parameters
$Q=P\cos(2\theta)$ and $U=P\sin(2\theta)$ as a function of the orbital
phase, calculated with the ephemeris $49000.0 + 13.6093 E$, with the
period from Mathieu (1994) and an arbitrary starting
point. Alencar et al. (2003) have found a periastron time of passage of
46654.3634$\pm$0.0086, which puts the periastron at around phase 0.64 in
our data. Observations are represented as first and second harmonics
of $\lambda=2\pi\phi$, where $\phi$ is the orbital phase, $0 < \phi < 1$
and the solid lines are fits made to the following equations:\\
\begin{eqnarray}
Q &=& q_0 + q_1 \cos \lambda + q_2 \sin \lambda + q_3 \cos 2\lambda +
q_4 \sin 2\lambda, \label{p3-eq-qfit}\\
U &=& u_0 + u_1 \cos \lambda + u_2 \sin \lambda + u_3 \cos 2\lambda +
u_4 \sin 2\lambda. \label{p3-eq-ufit}
\end{eqnarray}
The coefficients of this fit are then used to find the orbital
inclination according to the BME formalism (Brown, McLean \& Emslie 1978):\\
\begin{eqnarray}
\left[ \frac{1-\cos i}{1+\cos i} \right]^2 &=& \frac{(u_1+q_2)^2 +
(u_2-q_1)^2}{(u_2+q_1)^2 + (u_1-q_2)^2} \label{EQ-iO1-p3},\\
\left[ \frac{1-\cos i}{1+\cos i} \right]^4 &=& \frac{(u_3+q_4)^2 +
(u_4-q_3)^2}{(u_4+q_3)^2 + (u_3-q_4)^2} \label{EQ-iO2-p3}.
\end{eqnarray}

The polarization $P$ varies by 0.9\% in $U$, 1.0\% in $B$, and 0.8\% in
$V$. The position angle $\theta$ variations decrease from 47\arcdeg\ in
$U$, to 30\arcdeg\ in $B$, and to 22\arcdeg\ in $V$. Variations are then
more apparent in the blue (3550\AA\ $-$ 4300\AA) than in the yellow
(5250\AA) part of the spectrum.  

Data taken during 12 consecutive nights show clear and regular
variations, with some scatter but less than for many of the other binaries
that were observed over longer periods of time (Papers~IV and V). This
might be an indication that observations gathered over a small interval
of time will show less scatter and clearer periodic variations. Data
taken by Serkowski (1969) over 450 days, or even by Drissen et
al. (1989) over 40 days do not show the regular variations presented
here.

There is a strong peak (or dip) seen in $P$ and $U$ between phases 0.4
and 0.5, just before periastron, seen in all 3 filters. This peak is not
clearly present in the Serkowski data, but might be present in the
Drissen et al. (1989) data, only defined by one observation. The Bastien
(1985) data do not cover the appropriate orbital phases. In addition to
these single-periodic variations (seen once per orbit), the fit
indicates the presence of a smaller secondary peak. Numerical
simulations have shown that for circular orbits, only double-periodic
variations are seen; as the eccentricity increases, single-periodic
variations become predominant (see Papers~I and II). The presence of the
single-periodic variations is qualitatively compatible with the high
eccentricity, $e=0.47$. The ratio of the amplitude in 1$\lambda$ over
the amplitude in 2$\lambda$ is 7.34 for $Q$, 1.14 for $U$, 1.09 for $P$,
and 4.15 for $\theta$, at 5250\AA. However, the single-periodic
variations for $Q$ and $\theta$ clearly dominate over the
double-periodic ones, and this is not seen in our simulations for
$e=0.5$ and a CB disk unless the inclination is much lower ($i\approx
20\arcdeg$). AK~Sco and NTTS~162814-2427 (Paper~V) are the only two
examples so far of very strong single-periodic variations. Theses two
binaries also have a very high eccentricity (0.47 and 0.48
respectively).

Other explanations exist for the predominance of the single-periodic
variations: stars of different masses and luminosities, an asymmetric
configuration (with respect to the orbital plane) of the CB disk, or the
presence of a CS disk. We know that the luminosity ratio of the two
components is 4 (Andersen et al. 1989) but this is not enough to explain
the morphology of the variations. CS disks (although small) are
possible, along with a CB disk.

The strong amplitude of the $P$ variations, of over 0.5\%, is compatible
with the fact that a cleared central region is not necessarily indicated
by the spectral energy distribution; we have shown in Paper~I that when
scatterers are located close to the stars, they produce more pronounced
variations.

\subsection{Polarization as a function of wavelength}
Polarization is almost always lower in the blue and higher in the yellow
part of the spectrum. The position angle also varies as a function of
the wavelength, with different dependencies on the wavelength. Bastien
(1985) noted the highly variable wavelength dependence. Hutchinson et
al. (1994) also present a $P(\lambda)$ curve.

\section{Simultaneous photometry and polarimetry}
The photometric data are presented in Table~\ref{Tab-phot} and
Figure~\ref{Fig-phot}. Variations are clearly present, and follow the
same time-dependence for the 3 wavelengths. As can be seen in
Figure~\ref{Fig-colormag}, AK~Sco is bluer when bright, and redder when
faint, with no color reversal (when a further decrease in brightness is
followed by bluer colors again), a behavior that had already been found
by Bibo \& Th\'e (1991) using over 100 observations spread over more than
3 years, and by Hutchinson et al. (1994). A coarse study of the
photometric data presented by Bibo \& Th\'e (1991), especially in their
figure~29, shows that the photometric behavior of AK~Sco is not clearly
periodic but shows rapid variations. 

Hutchinson et al. (1994) also noted that the $10\mu$m flux increases
when the star gets fainter, an observation that is attributed to a major
change in the outer dust distribution ($\sim 1$~AU), indicating that a
simple occultation model is inadequate.

The color-magnitude diagram $B-V$, $V$ exhibits a ratio of total to
selective absorption of $R=4.3$, higher than in normal interstellar
clouds where $R=3.1$. This had already been noted by Andersen et al. (1989)
and would be characteristic of circumstellar (as opposed to IS) matter
(Strom et al. 1972).

There is also a correlation between the photometry and the polarimetry:
the polarization is low when the star is blue, and higher when the star
is redder (see Figure~\ref{Fig-polcolor}). Photometry and polarimetry in
the $B$ band is presented in Figure~\ref{Fig-bdata}. The behavior is
that when the star gets fainter, it also gets redder and its
polarization increases. 

There is no evidence in our data for eclipses of one component of the
binary by the other, as was suspected by Andersen et al. (1989). An
inclination higher than 63\arcdeg\ would however make an explanation of
the correlated photometric and polarimetric variations easier.
 
The simultaneous photometric and polarimetric observations increase the
diagnostic value of the data. The interpretation is that a cloud of
circumstellar matter passes in front of the star, decreasing the amount
of direct, unpolarized light, and increasing the contribution of
scattered and polarized light. This explains the decrease in brightness,
the reddening, and the increase in polarization.

\section{Orbital inclination and other parameters}
As shown in previous papers, the results of the BME formalism will only
give reasonable results if the stochastic noise present in the
data does not exceed some limit. Since AK~Sco's data was taken over a
short period of time (within one orbit), the data suffer from less noise
than for our other binaries (see Papers~IV and V). Even with data with less
scatter, the noise (square root of the variance of the fit over the
amplitude of the variations; the amplitude comes from the maximum values
of the data and not of the fit) is still above the 10\% limit.


Therefore, we cannot confirm or infirm the 63\arcdeg\ inclination found
by Andersen et al. (1989).


Even though the results of the BME formalism might not be significant,
at least for the orbital inclination, they are presented in
Table~\ref{Tab-BME} for the 3 wavelengths, where ${\cal O}1$ and ${\cal
O}2$ refer to the coefficients of single- and double-periodic
variations. The inclination found increases with wavelength, but this
might not be significant considering the uncertainties on the results.

AK~Sco has relatively high values of $\tau_0 G$ and $\tau_0 H$ (1-2
$\times 10^{-3}$), a factor of $\sim$ 10 higher than the values for
other  binaries (of all types), and the highest values for our sample of PMS
binaries. The ratios $\tau_0 G / \tau_0 H = 1.4-1.6$ are typical
values. It is interesting to note that AK~Sco has clear evidence for a
CB disk whose central region is not necessarily totally empty. GW~Ori,
NTTS~162814-2427 and NTTS~162819-2423S, which also have massive CB disks
but with central holes, have $\tau_0 G$ and $\tau_0 H$ values typical of
the other PMS binaries. The values of $\tau_0 G$, $\tau_0 H$, and the ratio
are similar for the $B$ and $V$ bands, but are different in the $U$
band.

\section{Comparison with numerical simulations}
Since the orbital inclination and eccentricity are known, we tried to
reproduce the observed polarimetric variations with various geometries. 
Even with these two parameters set, it is difficult to reproduce the
data, in particular the large observed polarization amplitude.

Figure~\ref{Fig-simulCB} presents the results of a simulation with a CB
disk surrounding the binary, which is at the center of an evacuated
cavity (see Figure~1 in Paper~II for a sketch of that geometry). The
disk's radius is set to 1.0 and has a flatness of 25\% (the third axis
equals 0.25 the disk radius); the central spherical cavity has a radius
of 0.20; the orbital parameters are $i=63\arcdeg$, $e=0.47$, and orbital
radius 0.1; astronomical silicate grains of radii 0.1$\mu m$ were used,
with an optical depth $\tau=0.1$ in the equatorial plane ($\tau=0.037$
in the line of sight). The stars have equal masses (and luminosities).
Variations are double-periodic but do not have the strong
single-periodic variations seen in the observations, and the amplitude
is smaller (0.02\% compared to the observed 0.8\%). If the inclination
is not as low as $i=63\arcdeg$ (Andersen et al. 1989), our simulations
still can not reproduce the amplitude or morphology of the observed
variations. 

Since we had found earlier that CS disk geometries are more favorable to
high amplitude variations, we did a simulation with only a CS disk
around the primary (see Figure~5 in Paper~II for a sketch of the
geometry). Figure~\ref{Fig-simulCS} presents the polarimetric variations
resulting from such a geometry. The CS disk has axes of radii 0.2, 0.2,
and 0.1, and a spherical cavity of radius 0.05. The same grains and
orbital characteristics were used, except that the orbital radius was
set to 0.5. The optical depth in the line of sight is now 0.07. The
primary illuminating its disk from the inside only introduces noise in
the polarization with amplitude 0.02\%. The figure shows only the
polarization caused by the secondary illuminating the disk
externally. The variations show two strong peaks close to each other
that do not really reproduce the observations; the amplitude is also too
small (0.1\%) although the variations produced are more important than
for the CB disk case. The polarization is highest around phase zero,
when the primary and its disk are closest to the observer and the
secondary is behind.

Since it is known that electrons usually produce more polarization and
higher amplitude variations, we tried the CS disk geometry with
electrons. There were no change in the shape of the variations nor their
amplitude. If the optical depth is increased from 0.1 to 0.3, the
amplitude of the variations goes from 0.09\% to almost 0.12\%. If the
periastron angle is changed from 0\arcdeg\ to 90\arcdeg, variations go
from 0.09\% to 0.17\%.

The maximum amplitude that we can get with our single scattering code,
with electrons to maximize $\Delta P$, but still retaining axial
symmetry for the distribution of scatterers around the primary (the only
asymmetry is that of the binary) is $\approx 0.2$\%. There are two other
options that would probably increase the amplitude of the polarization
variations, a larger optical depth, and also having a non-axisymmetric
density distribution around one (or both) stars. For example, a stream
of material falling onto the star along a magnetic flux tube as in some
recent observations for AA~Tau (Bouvier et al. 2003) or in the
high-resolution numerical simulations of G\"unther \& Kley (2002), would have
scatterers close to the star, where the polarization per scatterer is
larger. We leave these two possibilities to future modelling efforts.

\section{Summary and conclusions}
We have presented simultaneous photometric and polarimetric observations
of a PMS spectroscopic binary, AK~Sco. The mean observed polarization
($N$=27) is 0.70\% at 132\arcdeg\ at 3550\AA, 0.84\% at 132\arcdeg\ at
4300\AA, and 1.0\% at 133\arcdeg\ at 5250\AA. If an estimate of the
interstellar polarization is removed, the intrinsic polarization for
AK~Sco is still significant, 0.70\% at 108\arcdeg\ at 3550\AA, 0.83\% at
109\arcdeg\ at 4300\AA, and 0.94\% at 112\arcdeg\ at 5250\AA. This
indicates the presence of circumstellar matter distributed in an
asymmetric geometry (like a flattened envelope or disk).

AK~Sco's polarization cannot be of interstellar origin only because (1)
the position angles for the interstellar and observed polarizations are
different, (2) AK~Sco's polarization $P$ is variable, (3) its
position angle also varies, as a function of time and wavelength. 

The polarization is clearly variable at all wavelengths, on time scales
of hours, days, months, and years, which is a typical behavior for PMS
stars. The data were obtained on 12 consecutive nights, almost covering
the 13.6 day period. Regular variations are seen and have amplitudes of
$\Delta P \approx 1.0$\%, $\Delta \theta \approx 30^\circ$, the highest
seen so far for PMS binaries. Those variations are more apparent in the
blue (3550$-$4300\AA) than in the yellow part of the spectrum (5250\AA)
and are compatible with the presence of circumstellar matter located
close to the stars.

The variations are not simply double-periodic (as produced by a simple
model of 2 equal mass stars in a circular orbit, at the center of an
axisymmetric circumbinary envelope made of electrons), but include
single-periodic variations. The presence of single-periodic variations
could be due to non equal mass stars, the presence of dust grains, an
asymmetric configuration of the circumstellar or circumbinary material,
or the eccentricity of the orbit ($e=0.469$).

The polarimetric data show some scatter but less than for many
of the other PMS binaries that were observed over longer periods of time
(Papers~IV and V). We believe this indicates that observations of binaries
gathered over a small interval of time compared to the time scale
characteristic of modifications in the CS environment will suffer less
from epoch-to-epoch variations in the average polarization, and
therefore show less scatter and clearer periodic variations. 

The polarization varies as a function of wavelength, and is almost
always lower in the blue and higher in the red.  The position angle also
varies as a function of the wavelength, with different dependencies on
the wavelength. Both AK~Sco's polarization level and position angle vary
as a function of time and wavelength.

Large amplitude variations are also seen in photometry, with typical
amplitudes of 0.5~mag in $V$, 0.12~mag in $B-V$ and 0.20~mag in
$U-B$. These photometric variations are also correlated to the
polarimetric ones: usually, when the star is bright, its polarization is
small, and when the polarization is large, the star is fainter and
redder. These photometric and polarimetric variations are compatible
with occultation by a CS dust cloud in orbit in a dusty circumbinary
envelope. The color-magnitude diagram $B-V$, $V$ exhibits a ratio of
total to selective absorption of $R=4.3$, higher than in normal
interstellar clouds where $R=3.1$, and characteristic of circumstellar
(as opposed to interstellar) matter. There is no evidence for eclipses
of one component of the binary by the other, in both the photometric and
the polarimetric data.

Since numerical simulations produce high-amplitude variations for
circumstellar disks rather than for circumbinary disks, this suggests
the presence of CS (and maybe also CB) disks, although the short period
does not allow large CS disks. Alternatively, this indicates that there
is matter close to the stars, for example in a CB disk whose central
cavity is not completely evacuated, or in a stream from the CB disk to
the star.

A detailed analysis of the polarimetric data shows that the BME
formalism, which could be used to find the orbital inclination, is not
able to recover that information. Even though an inclination higher than
63\arcdeg would help explain the anticorrelation between the star's
brightness and its polarization if a cloud of dust passes in the line of
sight, the BME inclination between 80 and 90\arcdeg cannot be regarded
as a meaningful result. Therefore, we cannot confirm the
orbital inclinations found by Andersen et al. (1989) or Alencar et
al. (2003), 63\arcdeg, and 65\arcdeg\ $< i <$ 70\arcdeg\ respectively.

The BME results concerning the moments of the distribution are
interesting. AK~Sco has values of $\tau_0 G$ and $\tau_0 H$ a factor 10
higher than for other binaries. 

There are many free parameters, and even with the known inclination and
eccentricity, our numerical simulations could not reproduce the
observations. The maximum polarization amplitude we can obtain with a
single scattering model assuming an axially symmetric density
distribution of scatterers around the center of mass (CB envelope) or
around a single star (CS disk), is 0.20\%. Two avenues to explore
further are: take into account multiple scattering, and include
non-axisymmetric density distributions such as a stream of material
falling onto a component of the binary (or both).

\acknowledgments We thank the Conseil de Recherche en Sciences Naturelles et
G\'enie of Canada for supporting this research. The authors thank
A.~Le~Van~Suu for obtaining part of the photometric observations.


\newpage

\figcaption[Manset6.fig01.ps]{Map of the interstellar polarization in the
vicinity of AK~Sco. Polarization data are from the Heiles (2000)
catalog. The computed average interstellar polarization for AK~Sco is
shown at the position of the hexagon. See text for
details. \label{Fig-PolIS_AKSco}}

\figcaption[Manset6.fig02.ps]{Polarimetric observations of AK~Sco in the
3550\AA\ filter. The ephemeris used for the orbital phase is $49000.0 +
13.6093 E$, with the period from Mathieu (1994). Periastron occurs at
46654.3634$\pm$0.0086 (Alencar et al. 2003) which corresponds to a phase
of 0.64 in this figure and the following ones. The open circles are
duplicates of filled circles and were added for clarity. A few
observations are clearly away from the average curve, in both
polarization and position angle, and are due to intrinsic
variations. The solid line is the Fourier fit made according to
Equations~\ref{p3-eq-qfit} and \ref{p3-eq-ufit}. \label{Fig-3550all}}

\figcaption[Manset6.fig03.ps]{Same as Figure~2, for the
wavelength 4300\AA. \label{Fig-4300all}}

\figcaption[Manset6.fig04.ps]{Same as Figure~2, for the
wavelength 5250\AA. \label{Fig-5250all}}

\figcaption[Manset6.fig05.ps]{Photometric observations of AK~Sco from
1982 February. \label{Fig-phot}}

\figcaption[Manset6.fig06.ps]{Color-magnitude plot of AK~Sco in $V$ and
$B$. The solid line is the fit to the data and indicates a ratio of
total to selective absorption of $R=4.3$, lower than in IS clouds
(dashed line). \label{Fig-colormag}} 

\figcaption[Manset6.fig07.ps]{Polarization in the $V$ band as a
function of the $B-V$ color: the polarization is generally lower when
the star is bluer. \label{Fig-polcolor}}

\figcaption[Manset6.fig08.ps]{Photometry and polarimetry of AK~Sco in the
$B$ band. \label{Fig-bdata}} 

\figcaption[Manset6.fig09.ps]{Numerical simulation for AK~Sco surrounded by a
circumbinary disk. See text for details on the
geometry. \label{Fig-simulCB}}

\figcaption[Manset6.fig10.ps]{Numerical simulation for AK~Sco surrounded by a
circumstellar disk. See text for details on the
geometry. \label{Fig-simulCS}}


\newpage

\begin{deluxetable}{cccccccc}
\tablewidth{0pt}
\tablecaption{Averaged magnitudes and colors for the photometric
comparison stars \label{Tab-compstars}} 
\tablehead{
\colhead{Star} & \colhead{$V$} & \colhead{$\sigma_{V}$} &
\colhead{$B-V$} & \colhead{$\sigma_{B-V}$} &
\colhead{$U-B$} & \colhead{$\sigma_{U-B}$} &
\colhead{$N$}} 
\startdata
HR~8331 & 8.460 & 0.019 & 0.705 & 0.007 & 0.151 & 0.015 & 28\\
HR~8351 & 7.371 & 0.011 & 1.209 & 0.005 & 1.248 & 0.020 & 29
\enddata
\end{deluxetable}

\newpage

\begin{deluxetable}{ccccc}
\tablewidth{0pt}
\tablecaption{Photometry of AK~Sco \label{Tab-phot}}
\tablehead{
\colhead{JD} & \colhead{Phase\tablenotemark{1}} & 
\colhead{$V$} & \colhead{$B-V$} & \colhead{$U-B$}\\
2400000.0$+$ & \colhead{} & 
\colhead{} & \colhead{} & \colhead{}}
\startdata
45013.812 & 0.098& 8.839 & 0.622 & 0.179 \\
45013.832 & 0.099& 8.847 & 0.605 & 0.153 \\ 
45013.846 & 0.100& 8.850 & 0.601 & 0.143 \\
45013.879 & 0.103& 8.861 & 0.592 & 0.110 \\
45015.847 & 0.247& 8.814 & 0.611 & 0.142 \\
45015.876 & 0.249& 8.814 & 0.611 & 0.153 \\
45016.838 & 0.320& 9.027 & 0.648 & 0.251 \\
45016.845 & 0.321& 9.027 & 0.648 & 0.252 \\
45016.873 & 0.323& 9.067 & 0.651 & 0.260 \\
45016.880 & 0.323& 9.072 & 0.660 & 0.254 \\
45017.803 & 0.391& 9.237 & 0.685 & 0.281 \\
45017.810 & 0.391& 9.223 & 0.690 & 0.274 \\
45017.845 & 0.394& 9.215 & 0.662 & 0.296 \\
45017.852 & 0.395& 9.209 & 0.666 & 0.285 \\
45017.873 & 0.396& 9.204 & 0.666 & 0.281 \\
45017.878 & 0.397& 9.211 & 0.664 & 0.280 \\
45018.817 & 0.466& 9.336 & 0.708 & 0.308 \\
45018.824 & 0.466& 9.336 & 0.694 & 0.300 \\
45018.851 & 0.468& 9.349 & 0.684 & 0.301 \\
45018.858 & 0.468& 9.347 & 0.703 & 0.289 \\
45018.878 & 0.470& 9.346 & 0.696 & 0.302 \\
45018.884 & 0.470& 9.347 & 0.705 & 0.302 \\
45019.810 & 0.538& 9.209 & 0.649 & 0.192 \\
45019.816 & 0.539& 9.206 & 0.650 & 0.210 \\
45019.827 & 0.540& 9.209 & 0.647 & 0.199 \\
45019.833 & 0.540& 9.224 & 0.639 & 0.193 \\
45019.851 & 0.541& 9.204 & 0.659 & 0.197 \\
45019.856 & 0.542& 9.209 & 0.653 & 0.202 \\
45019.876 & 0.543& 9.211 & 0.652 & 0.203 \\
45019.881 & 0.544& 9.214 & 0.653 & 0.206 \\
45020.812 & 0.612& 9.237 & 0.665 & 0.246 \\
45020.819 & 0.613& 9.228 & 0.675 & 0.245 \\
45020.838 & 0.614& 9.224 & 0.670 & 0.234 \\
45020.842 & 0.614& 9.224 & 0.660 & 0.251 \\
45020.846 & 0.615& 9.216 & 0.670 & 0.243 \\
45020.868 & 0.616& 9.200 & 0.672 & 0.247 \\
45020.874 & 0.617& 9.209 & 0.664 & 0.256 \\
45020.887 & 0.618& 9.202 & 0.662 & 0.255 \\
45020.891 & 0.618& 9.192 & 0.669 & 0.244 \\
45021.814 & 0.686& 9.099 & 0.662 & 0.260 \\
45021.820 & 0.686& 9.100 & 0.670 & 0.258 \\
45021.871 & 0.690& 9.099 & 0.671 & 0.269 \\
45021.877 & 0.690& 9.106 & 0.659 & 0.265 \\
45021.888 & 0.691& 9.101 & 0.667 & 0.264 \\
45021.893 & 0.692& 9.102 & 0.670 & 0.271 \\
45022.878 & 0.764& 9.084 & 0.645 & 0.216 \\
45022.883 & 0.764& 9.090 & 0.648 & 0.225 \\
45023.788 & 0.831& 9.280 & 0.646 & 0.201 \\
45023.794 & 0.831& 9.282 & 0.649 & 0.203 \\
45023.815 & 0.833& 9.274 & 0.647 & 0.202 \\
45023.821 & 0.833& 9.280 & 0.643 & 0.178 \\
45023.852 & 0.835& 9.294 & 0.656 & 0.194 \\
45023.857 & 0.836& 9.298 & 0.653 & 0.194 \\
45023.873 & 0.837& 9.295 & 0.634 & 0.184 \\
45023.878 & 0.837& 9.288 & 0.644 & 0.183 \\
45024.804 & 0.905& 9.090 & 0.661 & 0.219 \\
45024.810 & 0.906& 9.090 & 0.658 & 0.215 \\
45024.827 & 0.907& 9.087 & 0.659 & 0.219 \\
45024.833 & 0.907& 9.087 & 0.651 & 0.220 \\
45024.870 & 0.910& 9.103 & 0.654 & 0.227 \\
45024.875 & 0.911& 9.094 & 0.657 & 0.221
\tablenotetext{1}{Calculated with the ephemeris $49000.0
+ 13.6093 E$, with the period from Mathieu (1994).}
\enddata	
\end{deluxetable}

\newpage

\begin{deluxetable}{cccccc}
\tablewidth{0pt}
\tablecaption{Polarization data for AK~Sco, 3550\AA \label{Tab-data3550}}
\tablehead{
\colhead{JD} & \colhead{Phase\tablenotemark{1}} & 
\colhead{P} & \colhead{$\sigma(P)$} &
\colhead{$\theta$} & \colhead{$\sigma(\theta)$}\\
2400000.0$+$ & \colhead{} &
\colhead{(\%)} & \colhead{(\%)} &
\colhead{(\arcdeg)} & \colhead{(\arcdeg)}}
\startdata
45013.793 & 0.096 & 0.466 & 0.101 &   71.9&  6.2\\ 
45013.852 & 0.101 & 0.572 & 0.104 &  123.6&  5.2\\  
45013.887 & 0.103 & 0.449 & 0.109 &  119.1&  6.9\\  
45014.790 & 0.170 & 0.602 & 0.097 &  119.8&  4.6\\  
45014.868 & 0.175 & 0.372 & 0.111 &  128.1&  8.5\\  
45015.798 & 0.244 & 0.375 & 0.092 &  123.7&  7.0\\  
45015.873 & 0.249 & 0.380 & 0.107 &  117.2&  8.1\\  
45016.793 & 0.317 & 0.408 & 0.160 &  140.3& 11.2\\  
45016.850 & 0.321 & 0.780 & 0.099 &  128.8&  3.6\\  
45017.806 & 0.391 & 1.251 & 0.087 &  130.0&  1.9\\  
45017.859 & 0.395 & 1.146 & 0.068 &  130.5&  1.7\\  
45018.811 & 0.465 & 1.254 & 0.128 &  143.8&  2.9\\  
45018.871 & 0.469 & 1.195 & 0.103 &  155.3&  2.5\\  
45019.782 & 0.536 & 0.830 & 0.147 &  138.6&  5.1\\  
45019.836 & 0.540 & 0.954 & 0.145 &  139.5&  4.4\\  
45019.884 & 0.544 & 1.112 & 0.124 &  135.9&  3.2\\  
45020.782 & 0.610 & 0.686 & 0.126 &  126.6&  5.3\\  
45020.823 & 0.613 & 0.860 & 0.112 &  136.0&  3.7\\  
45020.856 & 0.615 & 0.766 & 0.104 &  131.8&  3.9\\  
45021.781 & 0.683 & 0.711 & 0.113 &  132.6&  4.6\\  
45021.824 & 0.686 & 0.513 & 0.090 &  131.4&  5.0\\  
45021.864 & 0.689 & 0.513 & 0.092 &  123.8&  5.1\\  
45022.890 & 0.765 & 0.487 & 0.098 &  108.4&  5.8\\  
45023.810 & 0.832 & 0.544 & 0.100 &  117.5&  5.3\\  
45023.884 & 0.838 & 0.500 & 0.095 &  126.0&  5.4\\  
45024.779 & 0.904 & 0.618 & 0.088 &  133.1&  4.1\\
45024.853 & 0.909 & 0.450 & 0.082 & 134.3 & 5.2
\tablenotetext{1}{Calculated with the ephemeris $49000.0
+ 13.6093 E$, with the period from Mathieu (1994).}
\enddata
\end{deluxetable}

\newpage

\begin{deluxetable}{cccccc}
\tablewidth{0pt}
\tablecaption{Polarization data for AK~Sco, 4300\AA \label{Tab-data4300}}
\tablehead{
\colhead{JD} & \colhead{Phase\tablenotemark{1}} & 
\colhead{P} & \colhead{$\sigma(P)$} &
\colhead{$\theta$} & \colhead{$\sigma(\theta)$}\\
2400000.0$+$ & \colhead{} &
\colhead{(\%)} & \colhead{(\%)} &
\colhead{(\arcdeg)}  & \colhead{(\arcdeg)}}
\startdata
45013.779 & 0.095 & 0.516 & 0.060 & 134.7 & 3.3\\ 
45013.848 & 0.100 & 0.375 & 0.082 & 137.3 & 6.3\\ 
45013.890 & 0.103 & 0.744 & 0.062 & 125.0 & 2.4\\ 
45014.779 & 0.169 & 0.619 & 0.083 & 124.0 & 3.8\\ 
45014.863 & 0.175 & 0.744 & 0.068 & 132.7 & 2.6\\ 
45015.792 & 0.243 & 0.632 & 0.064 & 121.6 & 2.9\\ 
45015.869 & 0.249 & 0.579 & 0.059 & 126.7 & 2.9\\ 
45016.794 & 0.317 & 0.836 & 0.087 & 131.3 & 2.9\\ 
45016.850 & 0.321 & 0.703 & 0.077 & 128.9 & 3.1\\ 
45017.785 & 0.390 & 1.122 & 0.062 & 131.5 & 1.6\\ 
45017.870 & 0.396 & 1.311 & 0.053 & 129.5 & 1.2\\ 
45018.804 & 0.465 & 1.367 & 0.060 & 136.6 & 1.3\\ 
45018.876 & 0.470 & 1.380 & 0.081 & 149.9 & 1.7\\ 
45019.777 & 0.536 & 1.167 & 0.082 & 138.4 & 2.0\\ 
45019.832 & 0.540 & 1.077 & 0.080 & 132.8 & 2.1\\ 
45019.887 & 0.544 & 1.061 & 0.077 & 134.9 & 2.1\\ 
45020.776 & 0.609 & 0.971 & 0.079 & 125.5 & 2.3\\ 
45020.819 & 0.613 & 0.960 & 0.077 & 126.5 & 2.3\\ 
45020.861 & 0.616 & 0.814 & 0.074 & 126.6 & 2.6\\ 
45021.776 & 0.683 & 0.812 & 0.080 & 132.2 & 2.8\\ 
45021.826 & 0.687 & 0.530 & 0.077 & 131.5 & 4.2\\ 
45021.867 & 0.690 & 0.670 & 0.062 & 130.4 & 2.7\\ 
45022.896 & 0.765 & 0.596 & 0.067 & 144.8 & 3.2\\ 
45023.809 & 0.832 & 0.800 & 0.074 & 126.1 & 2.7\\ 
45023.884 & 0.838 & 0.897 & 0.068 & 127.7 & 2.2\\ 
45024.777 & 0.903 & 0.548 & 0.080 & 119.8 & 4.2\\ 
45024.852 & 0.909 & 0.599 & 0.074 & 130.4 & 3.5 
\tablenotetext{1}{Calculated with the ephemeris $49000.0
+ 13.6093 E$, with the period from Mathieu (1994).}
\enddata
\end{deluxetable}

\newpage

\begin{deluxetable}{cccccc}
\tablewidth{0pt}
\tablecaption{Polarization data for AK~Sco, 5250\AA \label{Tab-data5250}}
\tablehead{
\colhead{JD} & \colhead{Phase\tablenotemark{1}} & 
\colhead{P} & \colhead{$\sigma(P)$} &
\colhead{$\theta$} & \colhead{$\sigma(\theta)$}\\
2400000.0$+$ & \colhead{} &
\colhead{(\%)} & \colhead{(\%)} &
\colhead{(\arcdeg)}  & \colhead{(\arcdeg)}}
\startdata
45013.786 & 0.096 & 0.757 & 0.098 & 129.0 & 3.7\\  
45013.844 & 0.100 & 0.735 & 0.093 & 131.3 & 3.6\\  
45013.882 & 0.103 & 0.760 & 0.067 & 125.0 & 2.5\\  
45014.784 & 0.169 & 0.758 & 0.107 & 128.7 & 4.0\\  
45014.854 & 0.174 & 0.760 & 0.075 & 131.5 & 2.8\\  
45015.805 & 0.244 & 0.814 & 0.091 & 121.4 & 3.2\\  
45015.878 & 0.250 & 0.780 & 0.075 & 143.0 & 2.8\\  
45016.788 & 0.316 & 0.834 & 0.106 & 133.3 & 3.6\\  
45016.842 & 0.320 & 0.991 & 0.062 & 131.4 & 1.8\\  
45017.794 & 0.390 & 1.487 & 0.069 & 134.2 & 1.3\\  
45017.878 & 0.397 & 1.310 & 0.056 & 132.6 & 1.2\\  
45018.818 & 0.466 & 1.274 & 0.092 & 141.3 & 2.1\\  
45018.880 & 0.470 & 1.254 & 0.098 & 143.0 & 2.2\\  
45019.785 & 0.537 & 1.070 & 0.127 & 134.3 & 3.4\\  
45019.840 & 0.541 & 1.468 & 0.111 & 130.9 & 2.2\\  
45019.890 & 0.544 & 1.327 & 0.108 & 135.0 & 2.3\\  
45020.771 & 0.609 & 1.168 & 0.081 & 131.8 & 1.9\\  
45020.814 & 0.612 & 0.931 & 0.096 & 135.6 & 2.9\\  
45020.865 & 0.616 & 1.150 & 0.094 & 129.9 & 2.3\\  
45021.786 & 0.684 & 0.803 & 0.116 & 137.5 & 4.1\\  
45021.828 & 0.687 & 0.818 & 0.110 & 123.4 & 3.9\\  
45021.869 & 0.690 & 0.818 & 0.101 & 133.0 & 3.5\\  
45022.898 & 0.765 & 0.683 & 0.074 & 140.5 & 3.1\\  
45023.810 & 0.832 & 1.017 & 0.123 & 126.1 & 3.5\\  
45023.883 & 0.838 & 1.143 & 0.111 & 127.5 & 2.8\\  
45024.778 & 0.903 & 0.818 & 0.086 & 129.9 & 3.0\\  
45024.852 & 0.909 & 0.853 & 0.106 & 128.8 & 3.6 
\tablenotetext{1}{Calculated with the ephemeris $49000.0
+ 13.6093 E$, with the period from Mathieu (1994).}
\enddata
\end{deluxetable}

\newpage

\begin{deluxetable}{lcccccc}
\tablewidth{0pt}
\tablecaption{Results of the variability tests for AK~Sco\label{Tab-Var}}
\tablehead{
\colhead{Wavelength} & \colhead{} & 
\colhead{$\sigma_{\rm sample}$} & \colhead{$\sigma_{\rm mean}$} &
\colhead{$Z \pm \sigma_{Z}$} & \colhead{$P{\chi^2}$} &
\colhead{$P{\chi^2}$}\\
\colhead{(\AA)} & \colhead{} & 
\colhead{} & \colhead{} &
\colhead{} & \colhead{$1\sigma$} &
\colhead{$1.5\sigma$}}
\startdata
3550 &$Q$& 0.2414 & 0.0195 & 2.27 0.14 & 1.00 & 1.00 \\
     &$U$& 0.3355 & 0.0195 & 3.40 0.14 & 1.00 & 1.00 \\
4300 &$Q$& 0.2062 & 0.0135 & 2.74 0.14 & 1.00 & 1.00 \\
     &$U$& 0.2687 & 0.0135 & 4.01 0.14 & 1.00 & 1.00 \\
5250 &$Q$& 0.1783 & 0.0169 & 1.94 0.14 & 1.00 & 0.98 \\
     &$U$& 0.2481 & 0.0169 & 3.01 0.14 & 1.00 & 1.00
\enddata
\end{deluxetable}

\pagebreak \clearpage

\begin{deluxetable}{lccccccc}
\tablewidth{0pt}
\tablecaption{Results of the BME analysis for AK~Sco\label{Tab-BME}}
\tablehead{
\colhead{Wavelength} & \colhead{$i({\cal O}1)$} & \colhead{$\sigma(i)$} & 
\colhead{$i({\cal O}2)$} & \colhead{$\sigma(i)$} &
\colhead{$\tau_0 G$} & \colhead{$\tau_0 H$} & \colhead{$\tau_0 G /
\tau_0 H$}\\
\colhead{(\AA)} & \colhead{(deg)} & \colhead{(deg)} & 
\colhead{(deg)} & \colhead{(deg)} &
\colhead{$\times 10^{-3}$} & \colhead{$\times 10^{-3}$} & \colhead{}}
\startdata
3550 & 85.6&4.5 & 81.2&3.7 & 1.81 
     & 2.54 & 1.40\\
4300 & 86.3&4.4 & 84.7&2.9 & 1.31 
     & 2.10 & 1.60\\
5250 & 89.6&5.9 & 91.6&3.6 & 1.25 
     & 2.05 & 1.63
\enddata
\end{deluxetable}

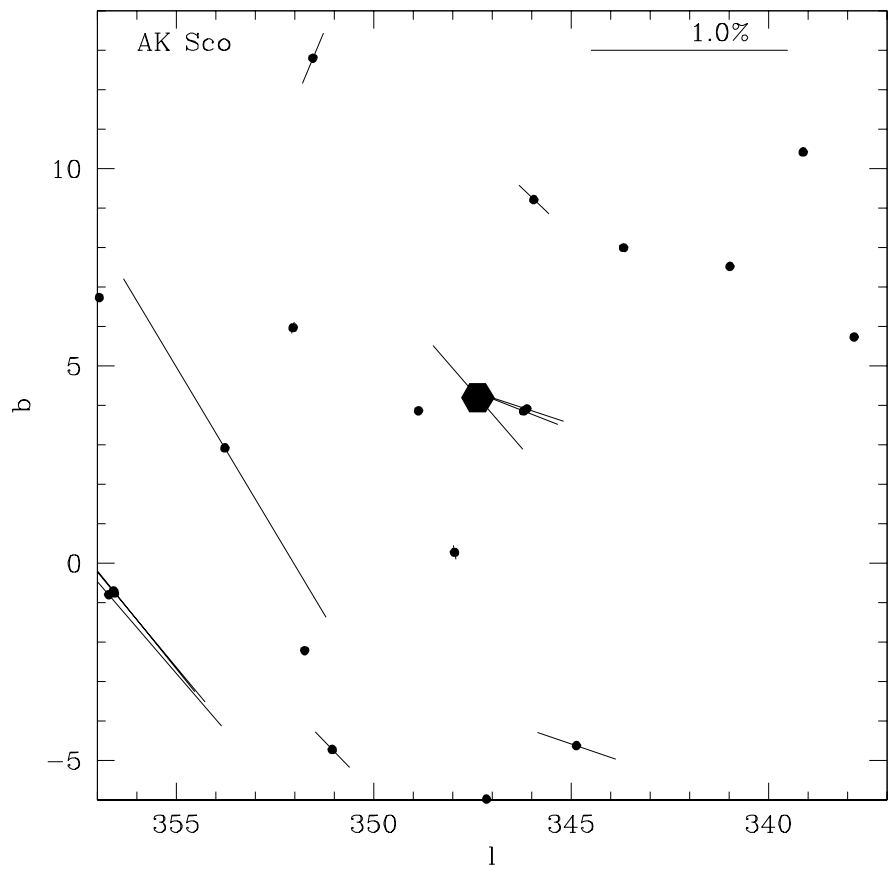
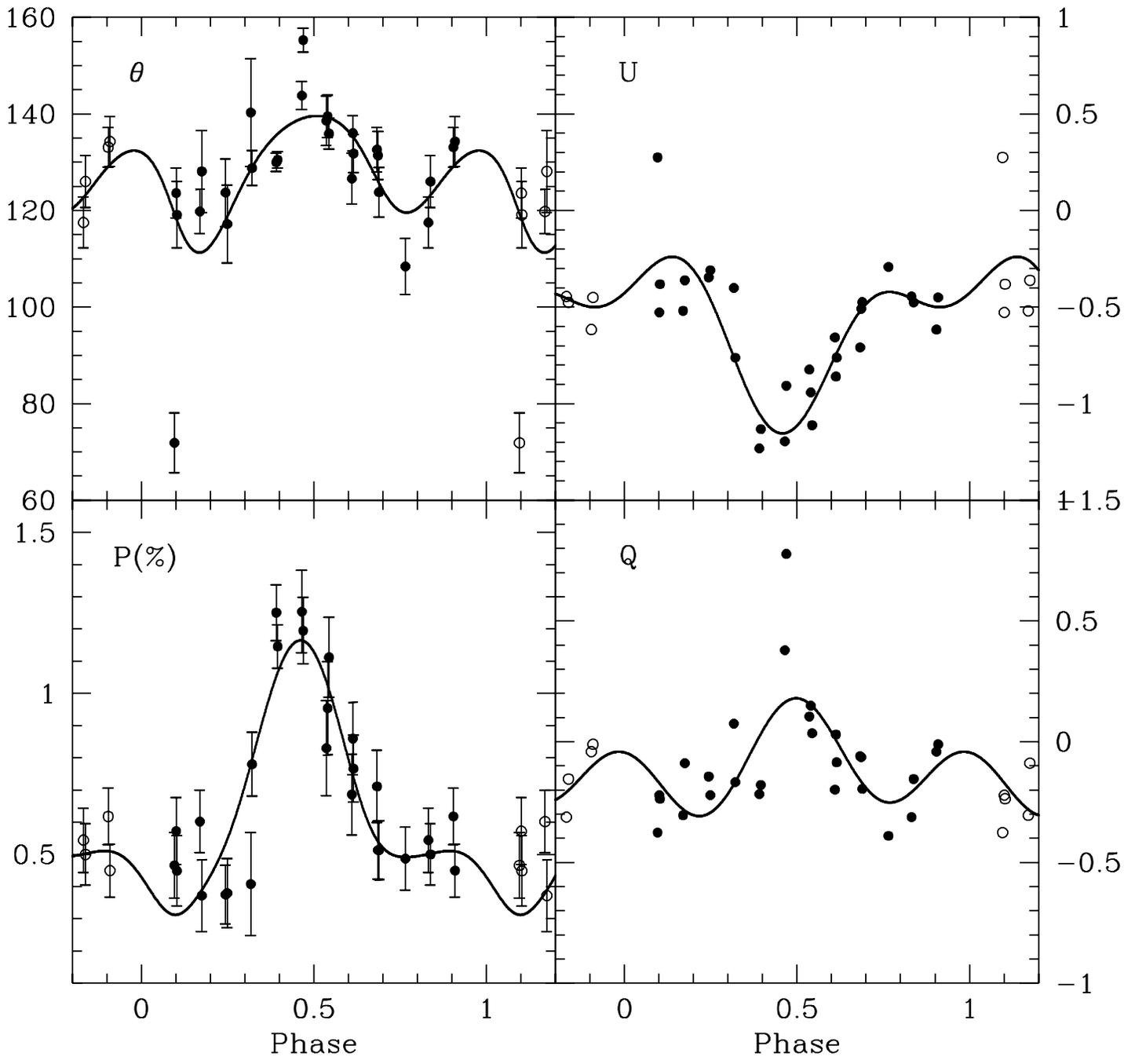
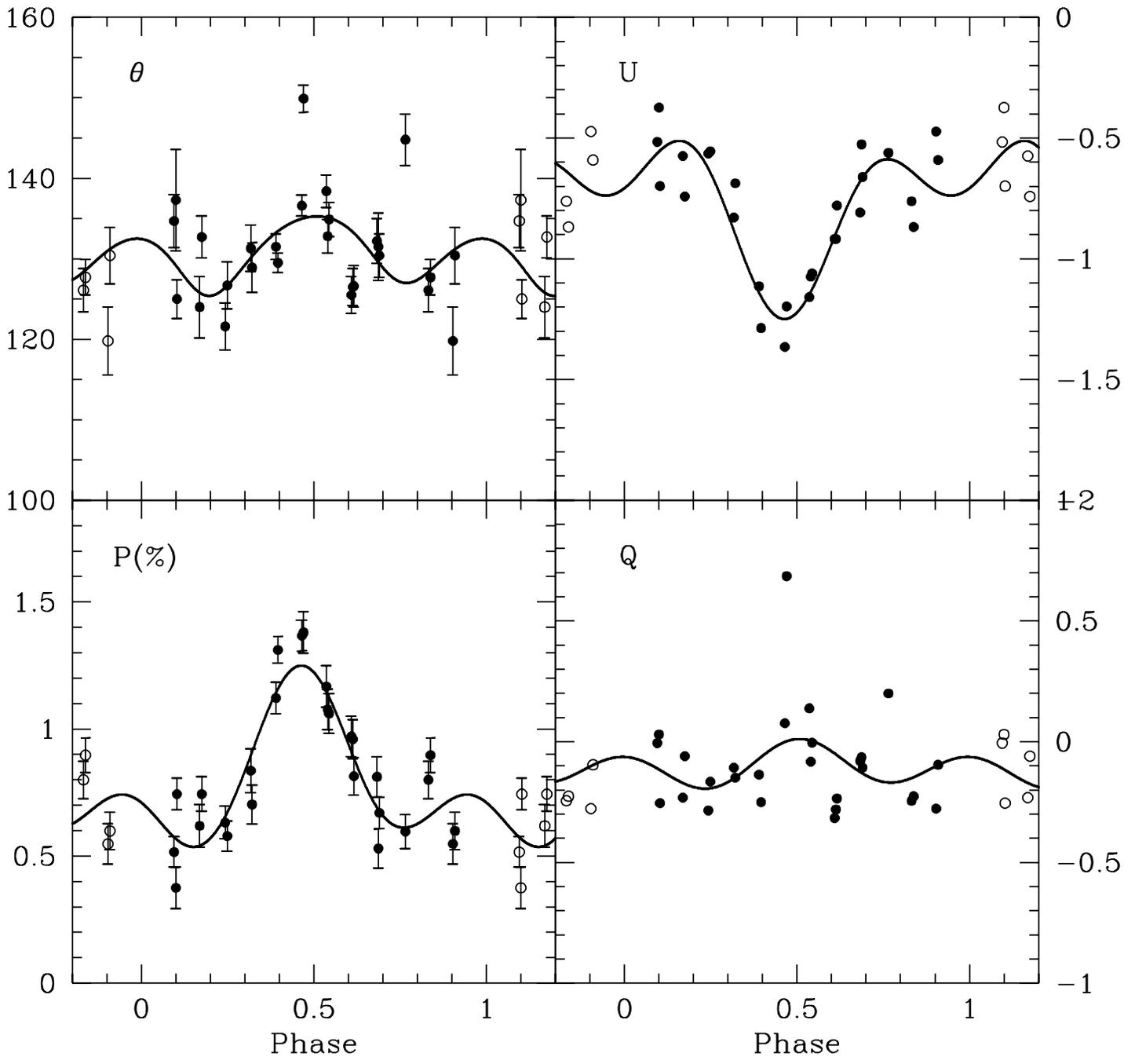
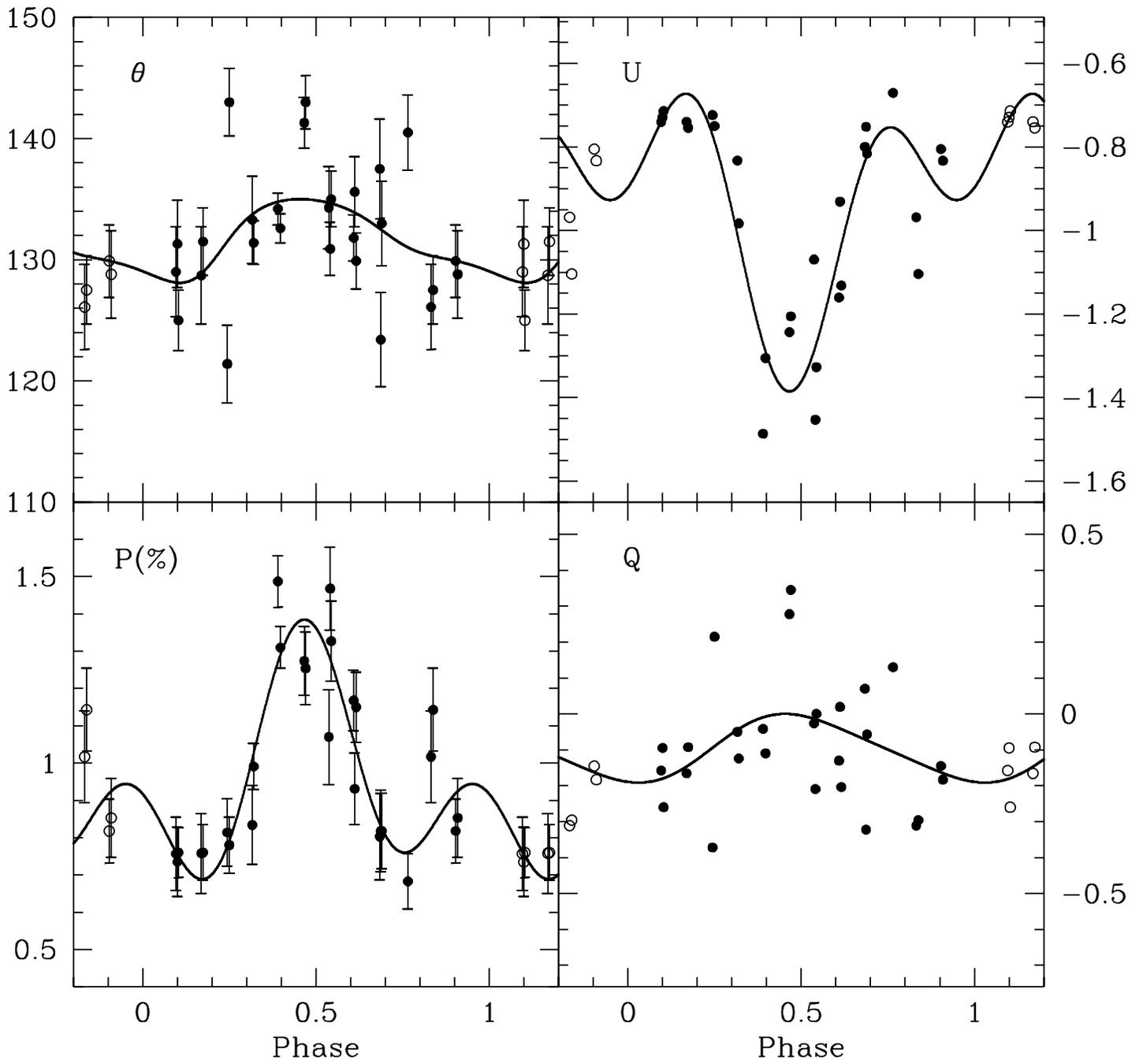
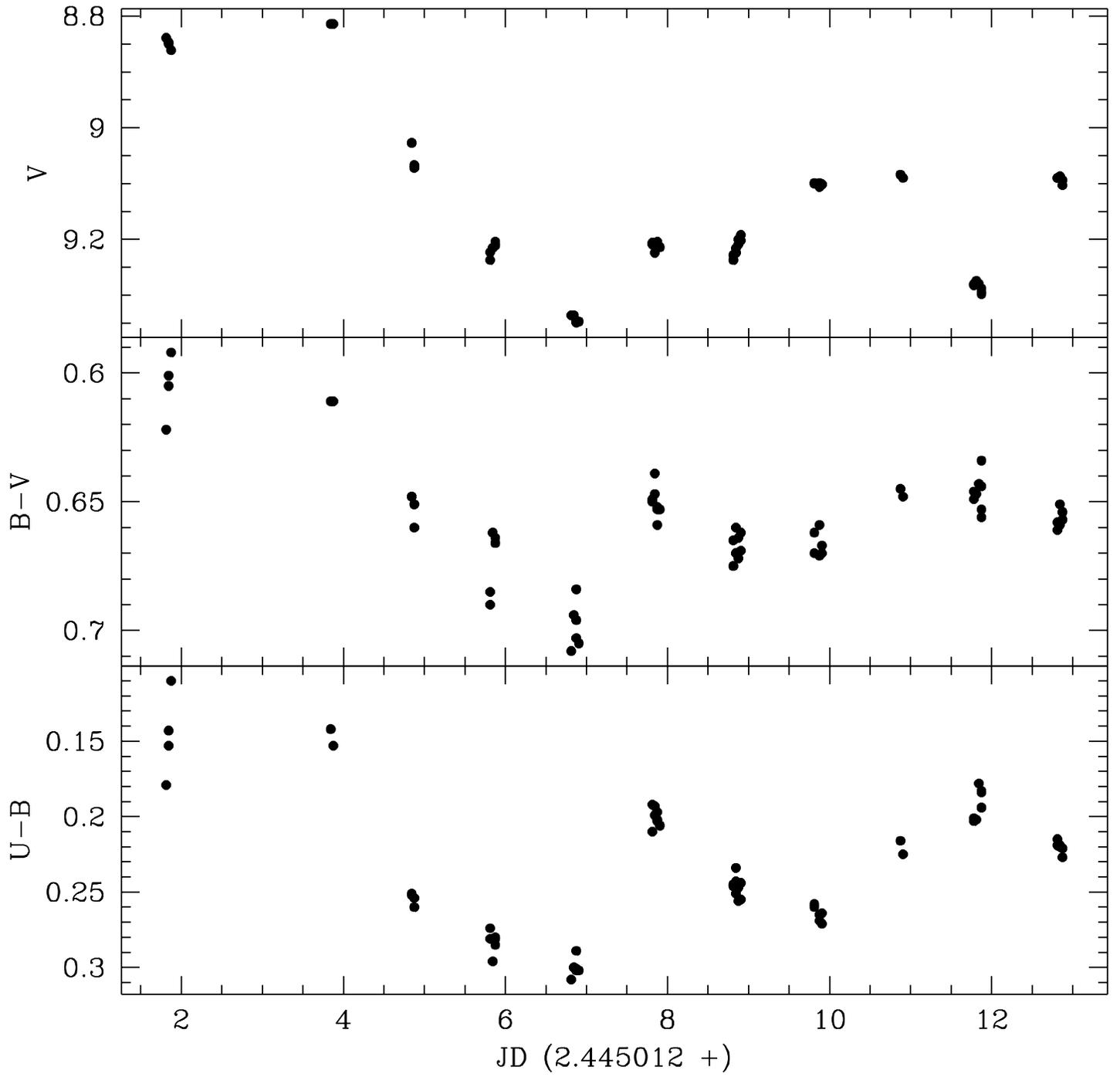
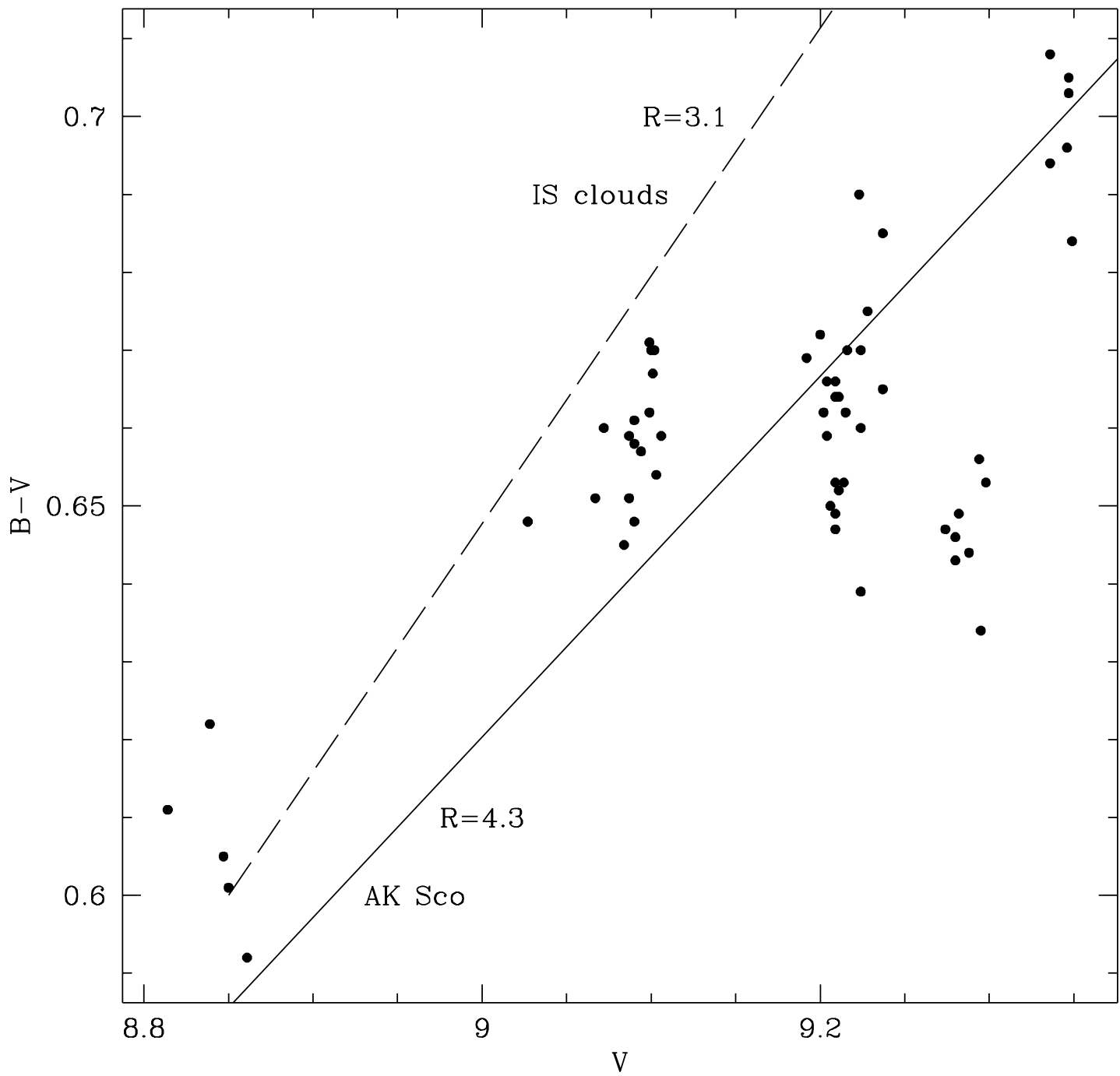
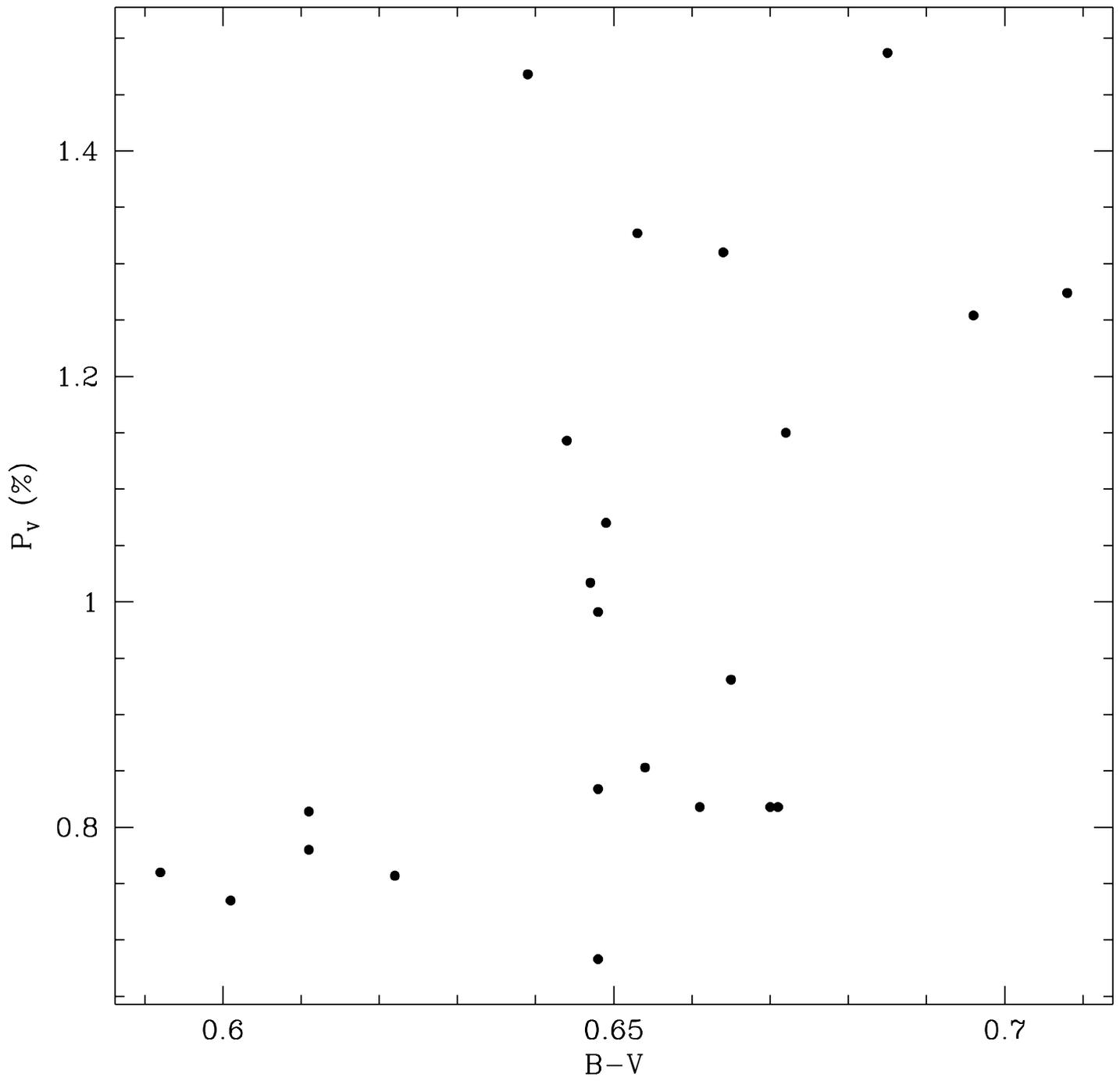
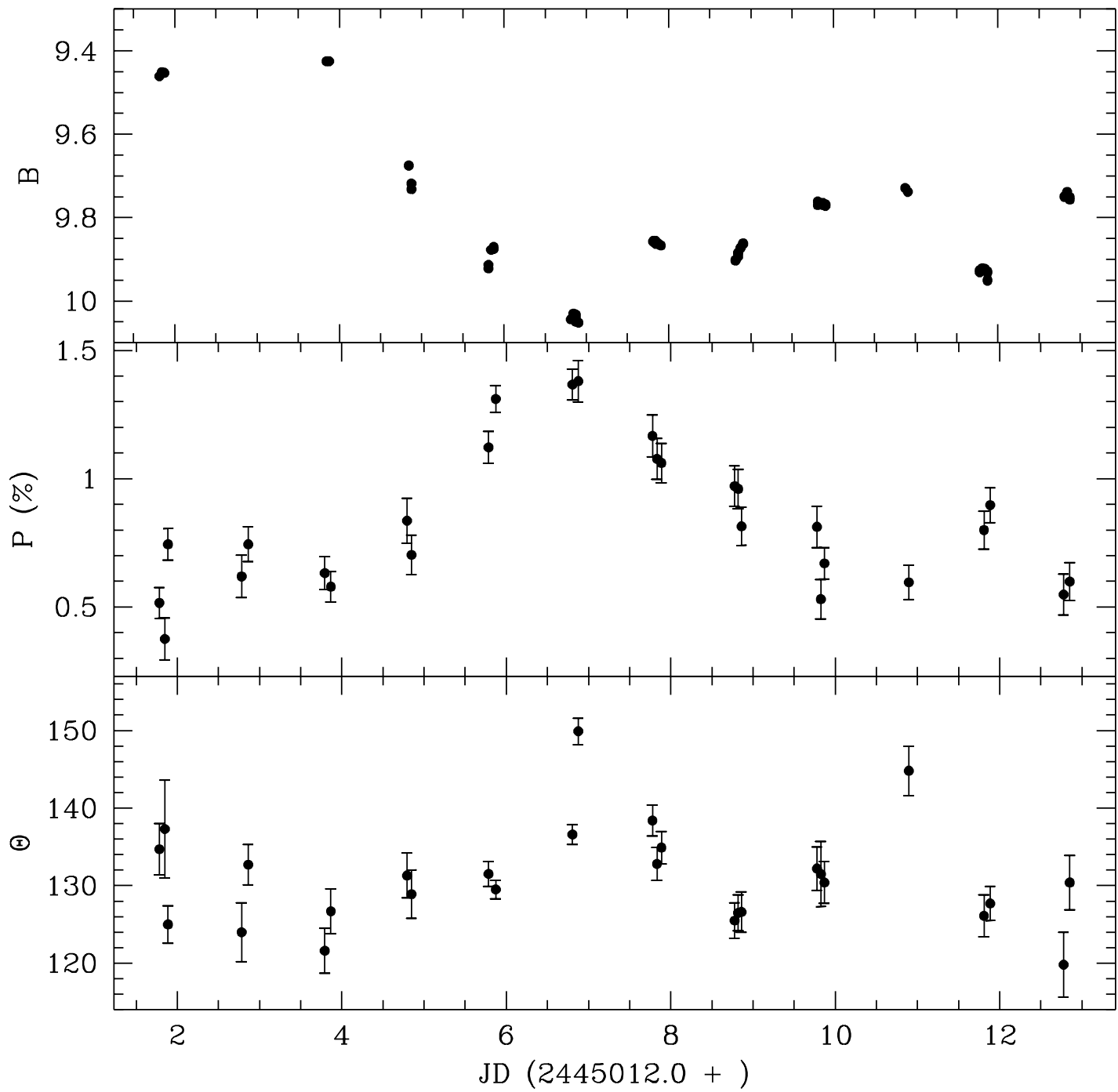
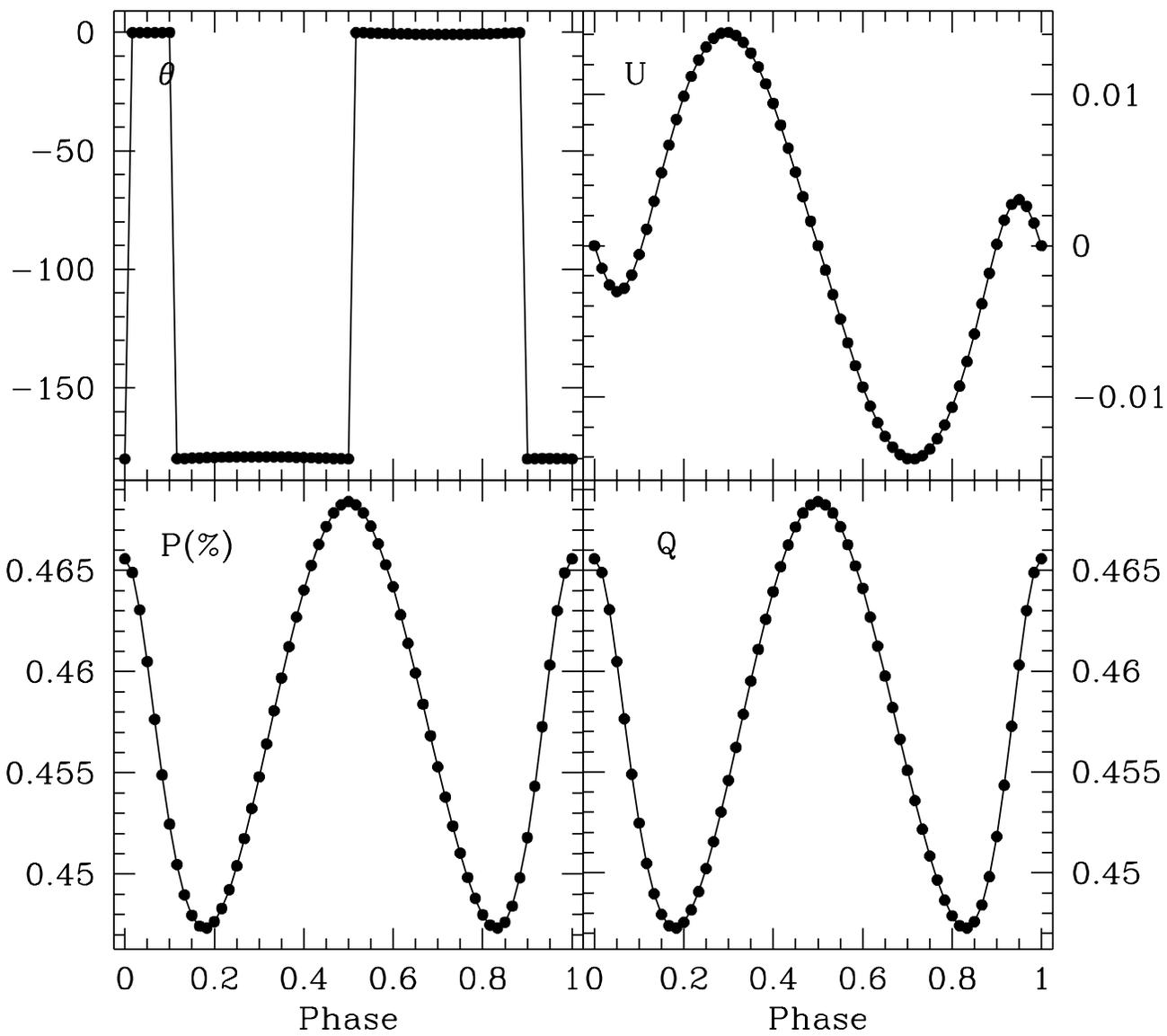
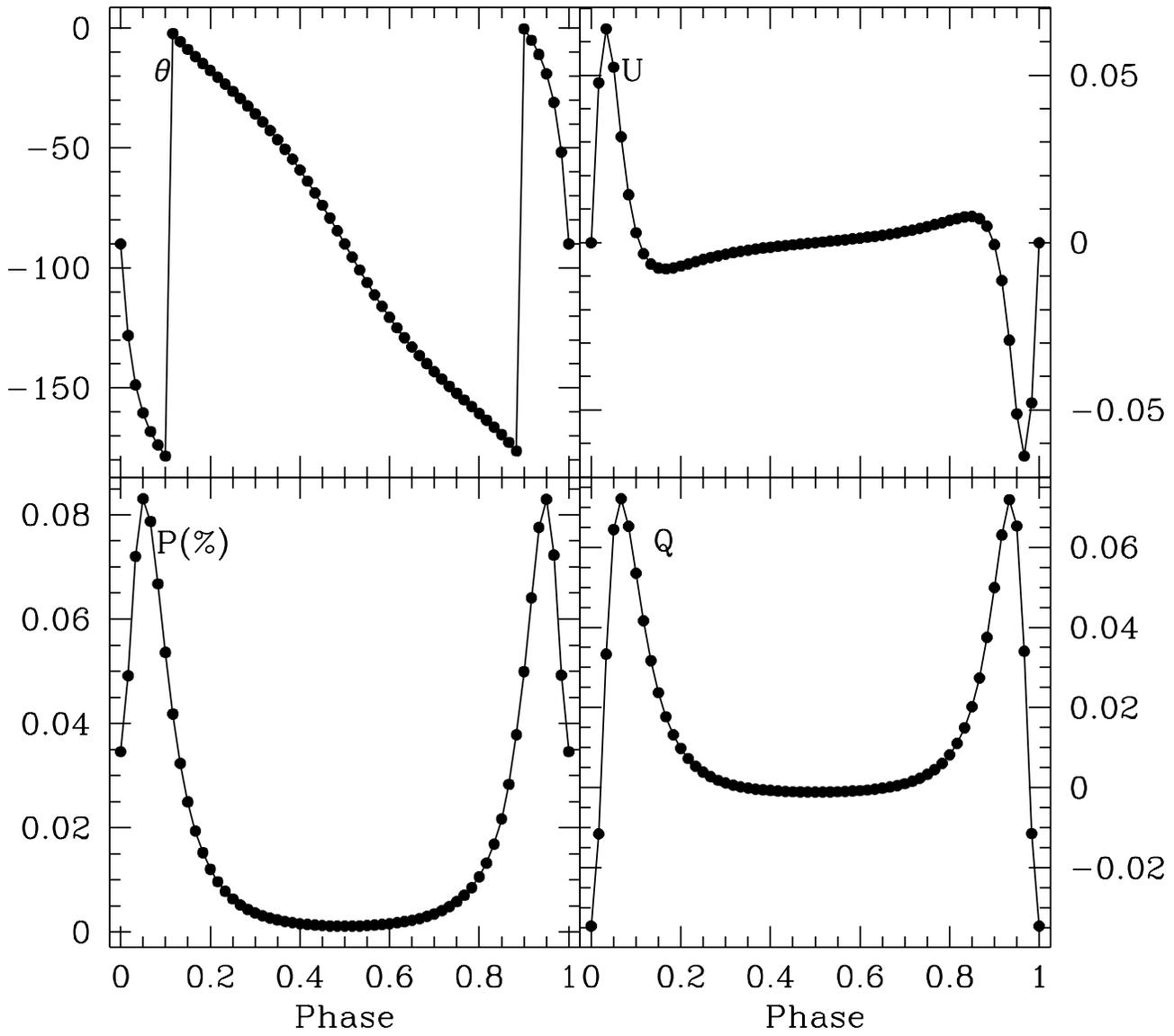
\end{document}